\RequirePackage{fix-cm}
\documentclass[smallextended]{svjour3}       
\smartqed  
\usepackage{amsmath}
\begin{document}

\title{Non-stationary thermal electromotive force generated by third sound}

\author{S.I. Shevchenko         \and
        A.M. Konstantinov
}

\authorrunning{S.I. Shevchenko, A.M. Konstantinov}

\institute{S.I. Shevchenko \at
              B.Verkin Institute for Low Temperature Physics and Engineering of the National Academy of Sciences of Ukraine, 47 Nauky Ave., Kharkov 61103, Ukraine \\
              Tel.: +380-93-5292460\\
              \email{shevchenko@ilt.kharkov.ua}
           \and
           A.M. Konstantinov \at
              B.Verkin Institute for Low Temperature Physics and Engineering of the National Academy of Sciences of Ukraine, 47 Nauky Ave., Kharkov 61103, Ukraine \\
              Tel.: +380-99-1864623\\
              \email{alexander.konstantinov2010@yandex.ua}
}

\date{Received: date / Accepted: date}

\maketitle

\begin{abstract}
It is predicted that oscillations of temperature during propagation of third sound in a thin superfluid film cause appearance of an alternating electric field in the surrounding space (a peculiar non-stationary thermoelectric effect). The magnitude of this field depends significantly on the substrate type and the method of its coating. It is shown that the differential thermal EMF (the ratio of electric potential amplitude to the film temperature amplitude) can exceed such one in metals and reach $10^{-4}$ V/K.
\end{abstract}

\section{Introduction}

\label{intro}
It is generally considered that passing of flow in a superfluid system (actually we consider only He II) is not accompanied by appearance of electric or magnetic fields in the surrounding space. In this article we will show that electric fields must appear during propagation of third sound \cite{1}, that is a surface wave with wavelength much exceeding the film thickness. In this case the normal component remains at rest (relative to the substrate) and the superfluid component oscillates parallel to the substrate. When the third sound is excited by creating a temperature difference between the ends of the film \cite{2}, local heating creates a gradient of superfluid density. As the result, the superfluid component of helium flows from the cold side of the film to the heated one. At non-stationary heating this process is accompanied by oscillations of film thickness. This, as will be shown in the article, causes oscillations of dipole moment of the film induced by the substrate and, as the result, appearance of oscillating electric field in the nearby space. The potential difference is proportional to the temperature gradient, i. e. a non-stationary thermoelectric effect must take place, that is impossible in normal systems.

\section{Electric field of a system "atom-solid"}

\label{sec:1}
The problem of polarization of an atom above a dielectric or metal substrate was considered by several authors \cite{3,4,5,6,7,8,9}. The general conclusion is that the dipole moment of an atom induced by the substrate decreases as the fourth power of the distance from the atom to the surface. However, we will be interested in the total electric field above the film, which contains contributions not only from dipole moments of the film atoms, but also of substrate atoms. In order to find this field we apply the approach \cite{10} used in \cite{11} for calculating the dipole moment of an atom.

Let an atom be located in vacuum at a distance $l$ from the surface of a solid with dielectric permeability $\varepsilon(\omega)$. We assume that $x$ and $y$ axes lie in the surface plane and $z$ axis is normal to the surface. The total Hamiltonian $H$ can be represented as a sum of the Hamiltonian of the atom and the solid without interaction between them $H_0$ and their interaction operator
\begin{equation}\label{1}
V_1=-d_iE_i(\mathbf{r}_a)-\frac{1}{6}Q_{ij}(\nabla_a)_jE_i(\mathbf{r}_a).
\end{equation}
Here $d_i$ and $Q_{ij}$ are operators of dipole and quadrupole moments of the atom respectively, ${\mathbf{r}_i=(0,0,l)}$ is the center of mass coordinate of the atom, $E_i(\mathbf{r})$ is the operator of the electric field created by the solid at the point $\mathbf{r}$. The total average electric field equals
\begin{equation}\label{2}
\langle\mathbf{E}^{tot}(\mathbf{r})\rangle\equiv\langle\psi\mid\mathbf{E}^{tot}(\mathbf{r})\mid\psi\rangle,
\end{equation}
where $\psi$ is the eigenfunction of the Hamiltonian $H$. As can be seen from subsequent calculations, it will be nonzero only if interactions of both dipole and quadrupole moments of the atom with the external field are taken into account. For calculating $\langle\mathbf{E}^{tot}(\mathbf{r})\rangle$ it is convenient to assume that there is a certain classic "external" dipole moment $\mathbf{d}_0$ at the point $\mathbf{r}$. As the result, the Hamiltonian obtains an addition
\begin{equation}\label{3}
V_2=-\mathbf{d}_0\mathbf{E}^{tot}(\mathbf{r})=-\mathbf{d}_0(\mathbf{E}(\mathbf{r})+\mathbf{E}_a(\mathbf{r})),
\end{equation}
where $\mathbf{E}_a(\mathbf{r})$ is the operator of the electric field created by the atom. If we introduce the $S$ operator (here and below $\hbar=c=1$)
\begin{equation}\label{4}
S=T_\tau\exp\left\{-i\int\limits_{-\infty}^{+\infty}{[V_1(t)+V_2(t)]dt}\right\},
\end{equation}
where $V_{1,2}(t)$ are operators $V_{1,2}$ in the interaction representation, $T_\tau$ is the chronological ordering operator, then the total energy of interaction of the electric field with the atom and the test dipole will be equal \cite{12}
\begin{equation}\label{5}
U=-\frac{1}{it}\langle S\rangle_0.
\end{equation}
Here $\langle S\rangle_0$ is the average of the $S$ operator by the ground state wave function of the isolated atom and the condensed medium. The average electric field is
\begin{equation}\label{6}
\langle\mathbf{E}^{tot}(\mathbf{r})\rangle=\langle\mathbf{E}(\mathbf{r})\rangle+\langle\mathbf{E}_a(\mathbf{r})\rangle=\left.-\frac{\partial U}{\partial\mathbf{d}_0}\right|_{\mathbf{d}_0=0}=\left.\frac{1}{it}\frac{\partial\langle S\rangle_0}{\partial\mathbf{d}_0}\right|_{\mathbf{d}_0=0}.
\end{equation}

We will consider a situation when the atom is situated at a distance from the solid that significantly exceeds the interatomic distance. In this case the interaction between the atom and the field $\mathbf{E}$ is small, and in order to find the average field $\langle\mathbf{E}^{tot}(\mathbf{r})\rangle$ it is possible to use the expansion of the $S$ operator (\ref{4}) by powers of $V_{1,2}(t)$. The first nonvanishing term in this expansion for the average field $\langle\mathbf{E}(\mathbf{r})\rangle$ created by the solid appears in the third order of perturbation theory:
\begin{equation}\label{7}
\langle E_i(\mathbf{r})\rangle=-\frac{1}{it}\frac{1}{3!}\int{dt_1dt_2dt_3dt_4}\langle T_\tau E_i(t_1,\mathbf{r})V_1(t_2)V_1(t_3)V_1(t_4)\rangle.
\end{equation}
This expression turns into zero if we do not consider the possibility of presence of a quadrupole moment in the atom, since in this case the integrand in (\ref{7}) is odd with respect either to the operator $\mathbf{E}$ or to the operator $\mathbf{d}$. Averaging of dipole and quadrupole moment operators in this expression is carried out over the ground state of the atom, and electric field operators are averaged over the ground state of the solid. Substituting (\ref{1}) here, we obtain the result
\begin{multline}\label{8}
\langle E_i(\mathbf{r})\rangle=-\frac{1}{it}\int{dt_1dt_2dt_3dt_4}D_{ij}^E(\mathbf{r}, \mathbf{r}_a, t_1, t_2)\Phi_{jklm}(t_2, t_3, t_4)\cdot\\(\nabla_a)_mD_{kl}^E(\mathbf{r}_a, \mathbf{r}_a, t_3, t_4).
\end{multline}
Here the Green function of a photon in the medium is
\begin{equation}\label{9}
D_{ij}^E(\mathbf{r}_1, \mathbf{r}_2, t_1-t_2)=\left\langle T_\tau E_i(\mathbf{r}_1,t_1)E_j(\mathbf{r}_2,t_2)\right\rangle,
\end{equation}
and a notation is introduced
\begin{equation}\label{10}
\Phi_{jklm}(t_1, t_2, t_3)=-\frac{1}{6}\left\langle T_\tau d_j(t_1)d_k(t_2)Q_{lm}(t_3)\right\rangle.
\end{equation}
The latter quantity appears in the expression for the average dipole moment of the atom above the solid \cite{6}
\begin{equation}\label{11}
\left\langle d_i\right\rangle=\frac{1}{t}\int{dt_1dt_2dt_3\Phi_{iklm}(t_1,t_2,t_3)(\nabla_a)_mD_{kl}^E(\mathbf{r}_a,\mathbf{r}_a,t_2,t_3)}.
\end{equation}
Turning to the function $D_{ij}=-i\left\langle T_\tau A_i(\mathbf{r}_1,t_1)A_j(\mathbf{r}_2,t_2)\right\rangle$ (where $\mathbf{A}$ is the vector potential operator), related to the Green function $D_{ij}^E$ by means of expression $D_{ij}^E(\mathbf{r}_1,\mathbf{r}_2,t)=i\partial^2/\partial t^2D_{ij}(\mathbf{r}_1,\mathbf{r}_2,t)$, and taking into account that $D_{ij}(\mathbf{r}_1,\mathbf{r}_2,\omega)=D_{ij}^R(\mathbf{r}_1,\mathbf{r}_2,|\omega|)$, where $D_{ij}^R(\mathbf{r}_1,\mathbf{r}_2,\omega)$ is the retarded Green function, after a number of transformations we find the average electric field created by the solid
\begin{equation}\label{12}
\langle E_i(\mathbf{r})\rangle=\left.-\left\langle d_z\right\rangle\left\{D_{iz}^R(\mathbf{r},\mathbf{r}_a,\omega)\omega^2\right\}\right|_{\omega=0}.
\end{equation}
As known \cite{13}, the retarded Green function satisfies an equation
\begin{equation}\label{13}
[\text{rot}_{lk}\text{rot}_{ki}-\omega^2\varepsilon(\omega,\mathbf{r})\delta_{il}]D_{ij}^R(\mathbf{r},\mathbf{r}_a,\omega)=-4\pi\delta_{lj}\delta(\mathbf{r}-\mathbf{r}_a),
\end{equation}
which is a consequence of Maxwell equations. Note that the second coordinate $\mathbf{r}_a$ and the second index $j$ do not participate in differential or algebraic operations performed on $D_{ij}^R(\mathbf{r},\mathbf{r}_a,\omega)$, i. e. they act only as external parameters. The function $D_{ij}^R(\mathbf{r},\mathbf{r}_a,\omega)$ must satisfy certain conditions at the solid boundary: functions $D_{ij}^R(\mathbf{r},\mathbf{r}_a,\omega)$ and $rot_{li}D_{ij}^R(\mathbf{r},\mathbf{r}_a,\omega)$ must be continuous (by the variable $\mathbf{r}$). It is possible to avoid necessity to solve the equation (\ref{13}), taking into account that this equation (at fixed values of $j$ and $\mathbf{r}_a$) coincides with the equation for the field $\mathbf{E}(\omega,\mathbf{r})$ created by a point dipole $d_l(\omega)=-\omega^{-2}\delta_{lj}$ located at the point $\mathbf{r}_a$. This field can be easily found using well-known electrostatic equations. However, we must bear in mind that we are now interested not in the total field $\left\langle\mathbf{E}^{tot}\right\rangle$, but only in the field from the solid. This field $\mathbf{E}(\omega,\mathbf{r})$ is caused by the image dipole
\begin{equation}\label{14}
\mathbf{d}_{im}(\omega)=\frac{\varepsilon(\omega)-1}{\varepsilon(\omega)+1}\left\{-d_x,-d_y,d_z\right\},
\end{equation}
emerging in the medium at the point $\mathbf{r}_{im}=(0,0,-l)$. As the result, we find the normal component of the average electric field (\ref{12})
\begin{equation}\label{15}
\left\langle {{E}_{z}}(\mathbf{r}) \right\rangle =\frac{\varepsilon (0)-1}{\varepsilon (0)+1}\cdot \frac{3{{\left( z+l \right)}^{2}}-{{\left| \mathbf{r}-{{\mathbf{r}}_{im}} \right|}^{2}}}{{{\left| \mathbf{r}-{{\mathbf{r}}_{im}} \right|}^{5}}}\left\langle {{d}_{z}} \right\rangle.
\end{equation}
The total electric field $\left\langle\mathbf{E}^{tot}(\mathbf{r})\right\rangle$ in (\ref{6}) is a sum of (\ref{15}) and the field created by the dipole $\langle d_z\rangle$. At large distances ($z\ll l$)
\begin{equation}\label{16}
\left\langle E_{z}^{tot}(\mathbf{r}) \right\rangle =\left[ 1+\frac{\varepsilon (0)-1}{\varepsilon (0)+1} \right]\frac{3{{z}^{2}}-{{r}^{2}}}{{{r}^{5}}}\left\langle {{d}_{z}} \right\rangle.
\end{equation}

This expression shows that the electric field of the system "atom-solid" at distances large compared to the distance from the atom to the surface is equivalent to the field of a dipole $p\equiv 2\varepsilon (0)\left\langle {{d}_{z}} \right\rangle /\left( \varepsilon (0)+1 \right)$ located directly on the substrate. Below we will regard this effective dipole as the atomic dipole.

\section{Flexoelectric effect}

Interaction of an atom with substrate is not the only possible mechanism of its polarization. Considering a helium film covering the substrate, we also need to take into account the dipole moments induced on the atom by neighboring film atoms. Interaction between two helium atoms causes redistribution of electron density in each of them. As the result, the pair of atoms acquires equal and oppositely directed dipole moments. Since the average distance ${R}$ between atoms in the helium film is much greater than the size of an atom and much less than characteristic wavelengths in the spectra of interacting atoms, the center of the electron density distribution in one atom turns out to be displaced towards the other one. The expression for the dipole moment induced by the other atom, has the form \cite{23}
\begin{equation}\label{41}
d_0=D_7\frac{ea_B^8}{R^7},
\end{equation}
where ${a_B}$ is the Bohr radius, ${D_7}\approx18.4$. The dipole moment of a certain atom in the film is obtained by adding dipole moments induced on this atom by all other film atoms. It turns out that, in the presence of inhomogeneity, atoms acquire nonzero dipole moments equal to \cite{8,9}
\begin{equation}\label{42}
\mathbf{d}_0=-\frac{4\pi D_7 ea_B^8}{9a^3}\nabla n_3.
\end{equation}
Here ${a}$ is the distance between helium atoms, ${n_3}$ is the three-dimensional density of atoms in the film. Thus, spatial inhomogeneity of the system generates a local dipole moment in it (flexoelectric effect).

The total dipole moment caused by inhomogeneity is compensated by dipole moments emerging in the surface layer of the film and in its layer adjacent to the substrate. Dipole moments in these layers significantly exceed the volume dipole moment (\ref{42}) and in the main approximation compensate each other. Electric potentials caused by these dipole moments also compensate each other in the same approximation. The next approximation yields corrections to the potentials of the same order as caused by dipole moments (\ref{42}). Therefore to estimate them it is sufficient to find the contribution of dipole moments (\ref{42}) into the potentials.

To estimate this contribution, we take into account that the density gradient ${\nabla n_3}$ in (\ref{42}) can be found from the balance condition of the pressure force ${\nabla p/\rho}$  (${\rho = mn_3}$ is the mass density of helium) and the Van der Waals force ${\mathbf{F}_\text{v}}$ acting on atoms from the substrate. Taking into account that ${\nabla p = (\partial p/\partial\rho)\nabla\rho=c_1^2\nabla\rho}$, where ${c_1}$ is the first sound speed in liquid helium, we find
\begin{equation}\label{43}
\nabla {{n}_{3}}=\frac{{{n}_{3}}}{{{c}_{1}}^{2}m}\cdot {{\mathbf{F}}_{\text{v}}}.
\end{equation}
If typical values of the plasma frequency ${\omega_p}$ (for a metal substrate) are much greater than the excitation frequencies of an atom ${\omega_a}$ and the distance from the atom to the substrate is not too large so that retardation effects are insignificant (this condition is fulfilled in the case of thin helium films considered here), then the Van der Waals force equals \cite{7}
\begin{equation}\label{44}
{{\mathbf{F}}_{\text{v}}}=\frac{\hbar {{\omega }_{a}}\alpha }{8{{z}^{4}}}\mathbf{\hat{z}}.
\end{equation}
Here ${z}$ is the distance between the atom and the substrate, ${\alpha}$ is static polarizability of a helium atom, $\mathbf{\hat{z}}$ is a unit vector directed normally to the medium surface. From (\ref{42}) - (\ref{44}) we can obtain
\begin{equation}\label{45}
{{\mathbf{d}}_{0}}=\frac{4\pi {{D}_{7}}e{{a}_{B}}^{8}{{n}_{3}}\hbar {{\omega }_{a}}\alpha }{72{{a}^{3}}m{{c}_{1}}^{2}}\frac{1}{{{z}^{4}}}\mathbf{\hat{z}}.
\end{equation}
Let us compare this dipole moment (\ref{45}) with the dipole moment (\ref{11}) induced on the atom by the substrate. The expression (\ref{11}) for the dipole moment $\left\langle {{d}_{z}} \right\rangle$ has been obtained in a different way in \cite{4}. If ${{{\omega }_{p}}\gg {{\omega }_{a}}}$, the dipole moment induced by the substrate is
\begin{equation}\label{46}
\left\langle \mathbf{d} \right\rangle =\frac{1}{8}\frac{{{\alpha }^{2}}}{e{{z}^{4}}}C\hbar {{\omega }_{a}}\frac{\varepsilon-1}{\varepsilon+1}\mathbf{\hat{z}},
\end{equation}
where ${C}$ is a numerical coefficient (for helium atoms ${C=9/4}$). The total dipole moment is obtained as a sum of (\ref{45}) and (\ref{46}). For the metal substrate ($\varepsilon\to \infty$) 
\begin{equation}\label{47}
{{\mathbf{d}}_{tot}}=\left\langle \mathbf{d} \right\rangle +{{\mathbf{d}}_{0}}=\left( 1+\frac{4\pi {{D}_{7}}{{e}^{2}}{{a}_{B}}^{8}{{n}_{3}}}{9C\alpha {{a}^{3}}m{{c}_{1}}^{2}} \right)\left\langle \mathbf{d} \right\rangle.
\end{equation}
Substituting numerical values into (\ref{47}) shows that the second term in the brackets is equal to 0.35, so the flexoelectric effect only slightly renormalizes the dipole moment of the atom induced by the substrate. Small renormalization of the dipole moment leads to small correction of the electric potential. This allows us to omit the correction caused by inhomogeneity.

\section{Polarization of a superfluid helium film}

If we suppose again that the excitation frequencies ${{\omega }_{a}}$ of the atom are much less than characteristic frequencies of oscillators in the medium ${{\omega }_{0}}$ (the plasma frequency ${{\omega }_{p}}$ in the case of a metal), the effective atomic dipole $p$ equals (see section 2 and equation (\ref{46}))
\begin{equation}\label{17}
\mathbf{p}\equiv \frac{2\varepsilon (0)\left\langle {{d}_{z}} \right\rangle }{\varepsilon (0)+1}\mathbf{\hat{z}}=\frac{1}{4}\frac{{{\alpha }^{2}}}{e{{z}^{4}}}C\hbar {{\omega }_{a}}\frac{\varepsilon (0)\left( \varepsilon (0)-1 \right)}{{{\left( \varepsilon (0)+1 \right)}^{2}}}\mathbf{\hat{z}}\equiv Ae{{a}_{B}}{{\left( \frac{{{a}_{B}}}{z} \right)}^{4}}\mathbf{\hat{z}}.
\end{equation}
Later on we will be interested in the electric field caused by presence of a helium film. For a helium-4 atom the polarizability $\alpha =2\cdot {{10}^{-25}}cm^3$ and the coefficient $C=9/4$. For a $^4\text{He}$ atom on a metal surface after substituting numerical values of the variables (taking into account that ${\hbar\omega_a=(1/2)e^2a_B^{-1}}$) we find $A\approx 0.5$.

Finding the expressions (\ref{16}) and (\ref{17}), we supposed that the helium atom is in vacuum. Since the dielectric constant of helium can be assumed equal to unity with sufficient accuracy, the results (\ref{16}) and (\ref{17}) remain valid also for atoms in helium film.
The dipole moment (\ref{17}) quickly decreases with increasing the distance between atoms and the substrate. Therefore polarization effects are significant only for thin helium films. This fact allows to assume that the film has a two-dimensional polarization which can be obtained by integrating the dipole moment $\mathbf{p}$ over the film thickness. Multiplying the result by the 3D density of helium atoms $n_3$, we find the total dipole moment per unit area
\begin{equation}\label{18}
{{\mathbf{P}}_{S}}=\frac{1}{3}Aea_{B}^{2}{{n}_{3}}\left[ {{\left( \frac{{{a}_{B}}}{a} \right)}^{3}}-{{\left( \frac{{{a}_{B}}}{h} \right)}^{3}} \right]\mathbf{\hat{z}}.
\end{equation}

When the film thickness $h$ and its density $n_3$ do not change in time, the substrate induces a constant dipole moment ${{\mathbf{P}}_{S}}$. Observation of electric fields caused by a constant dipole moment presents significant difficulties. It is easier to observe an addition to ${{\mathbf{P}}_{S}}$ varying in time and space. This addition, as can be seen from (\ref{18}), can be associated with variation of the density $n_3$ and the film height $h$. The first one takes place during propagation of fourth sound, the second one is caused by third sound. Oscillations of film height or density can be generated by creating a variable temperature difference between the film edges. The type of propagating sound oscillations is determined by balance of elastic forces and Van der Waals forces (see details in \cite{14}). We consider the case of third sound propagation. In this case the variable in (\ref{18}) is the film height $h$. Assuming the oscillations in the system are small, we write down the film height as $h={{h}_{0}}+{h}'(t,\mathbf{r})$, where ${h}'(t,\mathbf{r})\ll {{h}_{0}}$ (${{h}_{0}}$ is the equilibrium film height). The linear in ${h}'(t,\mathbf{r})$ addition to ${{\mathbf{P}}_{S}}$ equals
\begin{equation}\label{19}
{{\mathbf{P}'}_S}=Ae{{a}_{B}}{{\left( \frac{{{a}_{B}}}{{{h}_{0}}} \right)}^{4}}{{n}_{3}}{h}'(t,\mathbf{r})\mathbf{\hat{z}}\equiv P{h}'(t,\mathbf{r})\mathbf{\hat{z}}.
\end{equation}

So, let us consider propagation of third sound in a thin film whose temperature changes periodically at the edges. We assume that the film covers a flat substrate and lies in the $xOy$ plane, the sound wave propagates along the $x$ axis and the system is homogeneous in the $y$ direction. Temperature oscillations at the film boundaries will lead to temperature oscillations inside the film, causing superfluid component motion (the normal component in a thin film is fixed). This, in turn, leads to oscillations of the film height. Hence, magnitudes that describe the behavior of the film during third sound propagation are the temperature $T={{T}_{0}}+{T}'(t,x)$ (${{T}_{0}}$ is the equilibrium temperature), the superfluid component velocity ${{v}_{s}}={{v}_{s}}(t,x)$ and the film height $h={{h}_{0}}+{h}'(t,x)$.

Prior to writing down the equations for these variables, we mention that the behavior of the system depends on intensity of heat removal into the substrate, rate of evaporation of atoms from the film surface and vortex formation effects. It turns out that presence of vortices leads only to renormalization of third sound speed \cite{14,15,16} and their effect on the electric field caused by the third sound wave is weak even in the vicinity of the superfluid transition. Therefore we neglect the possibility of vortex formation in the system. Assuming the substrate is dielectric, we also neglect heat removal to the substrate. Taking into account evaporation and condensation of helium atoms, we obtain the equation describing the film behavior,
\begin{equation}\label{20}
\rho \frac{\partial {h}'}{\partial t}+{{h}_{0}}{{\rho }_{s}}\frac{\partial {{v}_{s}}}{\partial x}=-{{J}_{m}},
\end{equation}
\begin{equation}\label{21}
\rho s{{T}_{0}}\frac{\partial {h}'}{\partial t}+{{h}_{0}}\rho {{C}_{h}}\frac{\partial {T}' }{\partial t}=-{{s}_{g}}{{T}_{0}}{{J}_{m}}-{{J}_{Q}},
\end{equation}
\begin{equation}\label{22}
\frac{\partial {{v}_{s}}}{\partial t}=s\frac{\partial {T}'}{\partial x}-\frac{\gamma }{h_{0}^{4}}\frac{\partial {h}' }{\partial x}.
\end{equation}
Here ${{\rho }_{s}}$ is superfluid density, $\rho $ is the total film density, $s$ is specific entropy of helium, $\gamma $ is a parameter characterizing the intensity of Van der Waals forces acting on the film from the substrate, ${{C}_{h}}$ is heat capacity per unit mass of helium, ${{s}_{g}}$ is specific entropy of the vapor phase, ${{J}_{m}}$ is mass flow and ${{J}_{Q}}$ is heat flow from the film surface to the vapor phase.

Equation (\ref{20}) is a continuity equation with vaporization effects in its right-hand side. Equation (\ref{21}) expresses the entropy conservation law. The right-hand side of (\ref{21}) takes into account the entropy flow from the film to the vapor phase caused by mass flow (the first term) and by thermal conductivity of the vapor phase. The last equation (\ref{22}) is the equation of motion of the superfluid component. The right-hand side of this equation contains an expression for the chemical potential gradient, where Van der Waals forces acting on the film from the substrate (the second term) are considered. The expressions for ${{J}_{m}}$ and ${{J}_{Q}}$ have been obtained in \cite{17}
\begin{equation}\label{23}
{{J}_{m}}=\frac{4{{J}_{0}}}{{{k}_{B}}{{T}_{0}}}{{\left( 1-\frac{{{\rho }_{g}}}{\rho } \right)}^{-1}}\left[ \mu -{{\mu }_{g}}+\left( {{s}_{g}}-\frac{{{k}_{B}}}{2m} \right)\left( {T}'-{{{{T}'}}_{g}} \right) \right],
\end{equation}
\begin{equation}\label{24}
{{J}_{Q}}=-{{J}_{0}}\left[ \mu -{{\mu }_{g}}+\left( {{s}_{g}}-\frac{9{{k}_{B}}}{2m} \right)\left( {T}'-{{{{T}'}}_{g}} \right) \right].
\end{equation}
Here $\mu $ is the addition to the chemical potential of the liquid ($\mu =-s{T}'+(\gamma /{{h}_{0}}^{4}){h}'$), ${{\mu }_{g}}$ is the addition to the chemical potential of the vapor phase (${{\mu }_{g}}=-({{s}_{g}}+{{k}_{B}}/m){{{T}'}_{g}}+({{k}_{B}}T/m{{\rho }_{g}}){{{\rho }'}_{g}}$), ${{{\rho }'}_{g}}$ and ${{{T}'}_{g}}$ are oscillating parts of the vapor density and temperature respectively, ${{k}_{B}}$ is the Boltzmann constant, ${{J}_{0}}={{\rho }_{g}}{{\left( {{k}_{B}}{{T}_{0}}/8\pi m \right)}^{1/2}}$. The addition to the vapor density ${{{\rho }'}_{g}}$ present in the expressions for $\mu $ and ${{\mu }_{g}}$ can be eliminated using the motion equations of the vapor phase. As the result, (\ref{24}) yields
\begin{equation}\label{25}
{{J}_{Q}}=-{{J}_{0}}\left[ \frac{L}{{{T}_{0}}}{T}'+\frac{\gamma }{{{h}_{0}}^{4}}{h}'-\frac{9{{k}_{B}}}{2m}\left( {T}'-{{{{T}'}}_{g}} \right) \right].
\end{equation}
At not too low temperatures each of the terms in (\ref{25}), if not cancelled by other terms, leads to a large heat flow ${{J}_{Q}}$. In this case the right-hand side of (\ref{21}) is much greater than its left-hand side. To satisfy equation (\ref{21}), one should require fulfillment of the condition ${{J}_{Q}}=0$ \cite{17}. For non-saturated films ${T}'-{{{T}'}_{g}}\ll {T}'$ and the third term in (\ref{25}) can be omitted. Then it follows from (\ref{24}) that
\begin{equation}\label{26}
{h}'=-\frac{Lh_{0}^{4}{{{{T}'}}_{A}}}{\gamma {{T}_{0}}}{{e}^{i\left( \omega t-kx \right)}},
\end{equation}
where $L$ is latent heat of vaporization ($L={{T}_{0}}({{s}_{g}}-s)$), ${{{T}'}_{A}}$ is the amplitude of temperature oscillations. When third sound is created by temperature oscillations at the system boundaries, the magnitude of ${{{T}'}_{A}}$ is given, and after substituting ${h}'$ from (\ref{26}) into (\ref{19}) we find the connection between the polarization ${{\mathbf{{P}'}}_{S}}$ of the film and its temperature ${T}'$.

Considering (\ref{26}) and using equations (\ref{20}) - (\ref{22}), we can find the dispersion law of third sound. Assuming, as above, that ${T}'-{{{T}'}_{g}}\ll {T}'$, we obtain ${{\left( \omega /k \right)}^{2}}=c_{3}^{2}$, where the squared speed of third sound equals
\begin{equation}\label{27}
c_{3}^{2}=\frac{\gamma {{\rho }_{s}}}{\rho h_{0}^{3}}{{\left( 1+\frac{s{{T}_{0}}}{L} \right)}^{2}}.
\end{equation}
Note that the flows ${{J}_{m}}$ and ${{J}_{Q}}$ are proportional to the vapor phase density which decays exponentially with decreasing temperature. It has been shown in \cite{18,19} that at low temperatures the binding energy of a helium atom with the surface equals 7.15 K. Hence, at temperatures ${{T}_{0}}\ll 7.15 K$ right-hand sides of equations (\ref{20}) and (\ref{21}) will be exponentially small and can be omitted. Solving equations (\ref{20}) - (\ref{22}) with ${{J}_{m}}={{J}_{Q}}=0$, it is easy to find that now
\begin{equation}\label{28}
{h}'=-\frac{{{h}_{0}}{{C}_{h}}{{{{T}'}}_{A}}}{s}{{e}^{i\left( \omega t-kx \right)}},
\end{equation}
and the square of third sound velocity is
\begin{equation}\label{29}
{{c}_{3}}^{2}=\frac{{{\rho }_{s}}}{\rho }\left[ \frac{\gamma }{h_{0}^{3}}+\frac{{{s}^{2}}{{T}_{0}}}{{{C}_{h}}} \right].
\end{equation}

Note that the deviation of the film height ${h}'$ from its equilibrium value ${{h}_{0}}$ depends on ${{h}_{0}}$ in different ways. At low temperatures ${h}'\sim{{h}_{0}}$, but at high temperatures ${h}'\sim h_{0}^{4}$. The latter leads to a somewhat unexpected result: at high temperatures the addition to the equilibrium polarization ${{\mathbf{{P}'}}_{S}}$ does not depend on ${{h}_{0}}$. This result is valid while the temperature addition ${T}'$ creates a deviation ${h}'$ satisfying the condition ${h}'\ll {{h}_{0}}$.

\section{Thermal electromotive force generated by third sound}

Turning to calculation of electric fields generated in space by oscillating dipole moments of the film and the substrate, we bear in mind that a superfluid film crawls along a solid surface, covering the whole surface (if the solid temperature is lower than the superfluid transition point). Therefore in general the answer depends on the substrate shape. Below we limit ourselves with the case of a cylindrical vessel.

Let us consider a cylinder with internal radius ${{R}_{1}}$ and external radius ${{R}_{2}}$, assuming that ${{R}_{2}}+{{R}_{1}}\gg {{R}_{2}}-{{R}_{1}}$. Below we use the notation $R=\left( {{R}_{1}}+{{R}_{2}} \right)/2$. We choose a cylindrical coordinate system with its origin at the geometrical center of the cylinder and assume now that the $z$ axis is directed along the cylinder axis. Also we suppose that the cylinder radius $R$ and the coordinate of the observation point ${{z}_{0}}$ satisfy inequalities ${{L}_{z}}\gg R$ and ${{L}_{z}}\gg {{z}_{0}}$, where ${{L}_{z}}$ is the cylinder height.

Taking into account that third sound frequencies ${{\omega }}$ are significantly less than atomic frequencies, using the given polarization ${{\mathbf{{P}'}}_{S}}(\mathbf{r})$, one can find the electric field potential in the surrounding space using the known electrostatic formula
\begin{multline}\label{30}
\varphi ({{r}_{0}},{{z}_{0}})=\int{Rd\theta dz\,{{\mathbf{P}}_{S}}\cdot \frac{\partial }{\partial \mathbf{R}}\frac{1}{\left| \mathbf{R}-{{\mathbf{R}}_{0}} \right|}}= \\ \tilde{P}(t)\int\limits_{-\frac{{{L}_{z}}}{2}}^{\frac{{{L}_{z}}}{2}}{\int\limits_{-\pi }^{\pi }{\frac{R(R-{{r}_{0}}\cos \theta )\sin kz}{{{\left( {{R}^{2}}-2R{{r}_{0}}\cos \theta +r_{0}^{2}+{{\left( z-{{z}_{0}} \right)}^{2}} \right)}^{3/2}}}d\theta dz}}.
\end{multline}
Here ${{\mathbf{R}}_{0}}=\left( {{r}_{0}},{{\theta }_{0}},{{z}_{0}} \right)$ is the radius vector of the observation point. The expression for $\tilde{P}(t)$ has different forms at high and low temperatures. At high temperatures it follows from (\ref{19}) and (\ref{26}) that the two-dimensional polarization equals
\begin{equation}\label{31}
\tilde{P}(t)=\left( Ae{{a}_{B}}{{n}_{3}}{{\left( \frac{{{a}_{B}}}{{{h}_{0}}} \right)}^{4}}\frac{L{{h}_{0}}^{4}{{{{T}'}}_{A}}}{\gamma {{T}_{0}}} \right)\cos \omega t.
\end{equation}
At low temperatures we find from (\ref{19}) and (\ref{26})
\begin{equation}\label{32}
\tilde{P}(t)=\left( Ae{{a}_{B}}{{\left( \frac{{{a}_{B}}}{{{h}_{0}}} \right)}^{4}}\frac{{{n}_{3}}{{h}_{0}}{{C}_{h}}{{{{T}'}}_{A}}}{s} \right)\cos \omega t.
\end{equation}
The integral appearing in (\ref{30}) cannot be calculated at arbitrary ${{r}_{0}},{{z}_{0}}$. However, one can show that at ${{L}_{z}}\gg {{z}_{0}}$ the expression for $\varphi ({{r}_{0}},{{z}_{0}})$ is factorable and can be written in the form
\begin{equation}\label{33}
\varphi ({{r}_{0}},{{z}_{0}})=f({{r}_{0}})\sin k{{z}_{0}},
\end{equation}
where
\begin{equation}\label{34}
f({{r}_{0}})=\tilde{P}(t)\int\limits_{-\frac{{{L}_{z}}}{2}}^{\frac{{{L}_{z}}}{2}}{\int\limits_{-\pi }^{\pi }{\frac{R(R-{{r}_{0}}\cos \theta )\cos kzd\theta dz}{{{\left( {{R}^{2}}-2R{{r}_{0}}\cos \theta +{{r}_{0}}^{2}+{{z}^{2}} \right)}^{3/2}}}}}.
\end{equation}
Integration in (\ref{34}) is still not possible at arbitrary ${{r}_{0}}$, but the function $f({{r}_{0}})$ can be found using the following considerations. There are no free charges in the surrounding space, therefore, the potential $\varphi ({{r}_{0}},{{z}_{0}})$ must satisfy the Laplace equation $\nabla _{0}^{2}\varphi ({{r}_{0}},{{z}_{0}})=0$ (excluding the cylinder surface). Substituting (\ref{33}) into this equation shows that $f({{r}_{0}})$ has to satisfy the following equation
\begin{equation}\label{35}
\frac{{{d}^{2}}f}{d{{r}_{0}}^{2}}+\frac{1}{{{r}_{0}}}\frac{df}{d{{r}_{0}}}-{{k}^{2}}f=0.
\end{equation}
The solution of this equation, as is well known, has the form
\begin{equation}\label{36}
f({{r}_{0}})={{C}_{1}}{{I}_{0}}(k{{r}_{0}})+{{C}_{2}}{{K}_{0}}(k{{r}_{0}}).
\end{equation}
Here ${{I}_{0}}(k{{r}_{0}})$ and ${{K}_{0}}(k{{r}_{0}})$ are modified zero order Bessel functions of the first and the second kind respectively. Coefficients ${{C}_{1}}$ and ${{C}_{2}}$ change discontinuously when crossing the cylinder surface and can be found considering the following. At ${{r}_{0}}>R$ the function ${{I}_{0}}(k{{r}_{0}})$ increases exponentially, so, in this case the coefficient ${{C}_{1}}$ before ${{I}_{0}}(k{{r}_{0}})$ in (\ref{36}) should be taken zero. The coefficient ${{C}_{2}}$ before ${{K}_{0}}(k{{r}_{0}})$ can be found by calculating $f({{r}_{0}})$ at a certain value of ${{r}_{0}}$. It is convenient to do this at ${{r}_{0}}-R\ll R$. On the other hand, at ${{r}_{0}}<R$ we must take into account that ${{K}_{0}}(k{{r}_{0}})$ diverges at zero, hence, now we take ${{C}_{2}}=0$. To find ${{C}_{1}}$, it is again sufficient to find $f({{r}_{0}})$ at a certain particular value of ${{r}_{0}}$. The easiest way is to perform this at ${{r}_{0}}=0$. Finally, we obtain
\begin{equation}\label{37}
\varphi ({{r}_{0}},{{z}_{0}})=4\pi \tilde{P}(t)(kR)\sin k{{z}_{0}}\cdot \left\{ \begin{matrix}
   -{{I}_{1}}(kR){{K}_{0}}(k{{r}_{0}}),\ {{r}_{0}}>R  \\
   {{K}_{1}}(kR){{I}_{0}}(k{{r}_{0}}),\ \ \ {{r}_{0}}<R  \\
\end{matrix} \right..
\end{equation}

During the derivation of (\ref{37}) we assumed the inequality $k{{L}_{z}}\gg 1$ to be valid. This inequality leaves the magnitude of $kR$ arbitrary. The behavior of the potential $\varphi $ as a function of $k{{r}_{0}}$ depends significantly on $kR$. At $kR\ll 1$ and $k{{r}_{0}}\ll 1$ it follows from (\ref{37}) that the total electric field potential above the film equals
\begin{equation}\label{38}
\varphi ({{r}_{0}})=4\pi \tilde{P}(t)\sin k{{z}_{0}}\left\{ \begin{matrix}
   \left( {{\left( kR \right)}^{2}}/2 \right)\ln k{{r}_{0}},\ \quad {{r}_{0}}>R  \\
   \quad \quad \quad 1,\quad \ \quad \quad \quad {{r}_{0}}<R  \\
\end{matrix} \right..
\end{equation}
At $kR\ll 1$ and $k{{r}_{0}}\gg 1$
\begin{equation}\label{39}
\varphi ({{r}_{0}})=2\pi \tilde{P}(t){{\left( kR \right)}^{2}}\sqrt{\frac{\pi }{2k{{r}_{0}}}}\exp \left( -k{{r}_{0}} \right)\sin k{{z}_{0}}.
\end{equation}
At $kR\gg 1$ the potential $\varphi ({{r}_{0}})$ is not small if the observation point ${{r}_{0}}$ is in the vicinity of the cylinder surface. In this case
\begin{equation}\label{40}
\varphi ({{r}_{0}})=\text{sgn} \left( R-{{r}_{0}} \right)2\pi \tilde{P}(t)\exp \left( -k\left| R-{{r}_{0}} \right| \right)\sin k{{z}_{0}}.
\end{equation}
Note that expressions (\ref{38}) - (\ref{40}) are approximate (corrections of the order of $k{{r}_{0}}\ll 1$ have been omitted), therefore they will not satisfy the Laplace equation. For the exact expression (\ref{37}) we have $\nabla _{0}^{2}\varphi ({{r}_{0}},{{z}_{0}})=0$.

Thus, oscillations of temperature on the film edges cause oscillations of the film height. Height oscillations lead to oscillations of the dipole moment induced by the substrate and by interaction of atoms. These dipole moment oscillations generate electric fields in the surrounding space. As the result, we find out existence of relation between electric fields and temperature, i. e. a peculiar thermoelectric effect.

In order to find the connection between the temperature gradient in the film and the electric potential generated by this gradient, one should substitute (\ref{31}) at high temperatures and (\ref{32}) at low temperatures into (\ref{37}). As we can see, the magnitude of this potential depends significantly on the observation point. At $kR\gg 1$ maximum values of the potential $\varphi ({{r}_{0}})$ are reached near the cylinder surface, no matter if the film covers the cylinder inside or outside. At $kR\ll 1$, due to presence of the factor ${{\left( kR \right)}^{2}}\ll 1$, the electric potential amplitude is much greater if the film covers the cylinder from the inside.

Let us estimate the magnitude of the electric potential. For low temperatures we should use expression (\ref{32}). Entropy and heat capacity of at these conditions are determined by phonon gas, so ${{C}_{h}}/s=3$ (see \cite{20}). The electric potential value strongly depends on the equilibrium film thickness ${{h}_{0}}$. It follows from (\ref{32}) that at low temperatures $\varphi \sim h_{0}^{-3}$. The smallest equilibrium thickness of a film in which third sound has been observed \cite{21}, is two atomic layers. Assuming ${{h}_{0}}=2\cdot a\approx 6\cdot {{10}^{-8}}\text{ cm}$ (where $a$ is the interatomic distance) and ${{n}_{3}}=2\cdot {{10}^{22}}\text{ c}{{\text{m}}^{-3}}$, we find the maximum possible value of the electric potential amplitude $\left| \varphi  \right|={{{T}'}_{A}}\cdot 6\cdot {{10}^{-4}}\text{ V/K}$.

For high temperatures the estimate of the potential can be obtained using (\ref{31}) and (\ref{37}). As mentioned above, in this case the potential does not depend on the equilibrium film thickness. Assuming $L={{10}^{8}}\text{ erg/g}$, $\gamma =2.2\cdot {{10}^{-14}}\text{ erg}\cdot \text{c}{{\text{m}}^{3}}\text{/g}$ and ${{n}_{3}}=2\cdot {{10}^{22}}\text{ c}{{\text{m}}^{-3}}$, we find $\left| \varphi  \right|=2\cdot {{10}^{-4}}\text{ V}\cdot \left( {{{{T}'}}_{A}}/{{T}_{0}} \right)$. At ${{T}_{A}}^{\prime }\sim{{10}^{-3}}\text{ K}$ this estimate is valid up to thicknesses ${{h}_{0}}\approx 20\cdot a\approx 6\cdot {{10}^{-7}}\text{ cm}$.

The ratio $\left| \varphi  \right|/{{T}_{A}}^{\prime }$, as is well known, is called differential thermal EMF. This particular quantity is a characteristic of thermoelectric properties of a substance. For pure non-magnetic metals this quantity has the order of $10^{-6}-10^{-8}$ V/K (see e. g. \cite{22}). The cause of the large differential thermal e.m.f. is an anomalously large thermomechanical effect in superfluid helium. For the amplitude value of temperature assumed above (${{T}_{A}}^{\prime }\sim{{10}^{-3}}\text{ K}$) the thermoelectric potential $\varphi $ will be rather small ($\varphi <{{10}^{-7}}\text{ V}$).

In summary, we have shown in this article that two unique properties of superfluid systems, namely, ability to flow along a rugged surface without dissipation and an anomalously large thermomechanical effect lead to one more unique property, that is a non-stationary thermoelectric effect caused by third sound. This effect is an electrical analogue of the fountain effect.



\end{document}